\documentclass[12pt]{article}
\usepackage{graphics}
\usepackage{amsfonts}
\usepackage{amsmath}
\usepackage{amssymb}

 \let\b=\beta  \let\g=\gamma   \let\e=\varepsilon
  \let\h=\eta   \let\th=\theta  
\let\m=\mu    \let\n=\nu             \let\r=\rho

\let\O=\Omega

\def\\{\hfill\break} \let\==\equiv

\let\0=\noindent

\def\qed{\hfill\raise1pt\hbox{\vrule height5pt width5pt depth0pt}}

\def\be{\begin{equation}}
\def\ee{\end{equation}}
\def\bea{\begin{eqnarray}}\def\eea{\end{eqnarray}}

\begin{document}
%----------------------------------------------------------------

\markright{Hasen\"ohrl...}
%--------------------------------------------------------------------

\title{Fritz Hasen\"ohrl and $E=mc^2$}

\author {Stephen Boughn{\small\it\thanks{sboughn@haverford.edu}}
\\[2mm]
~ \it Haverford College, Haverford PA 19041}

\date{{\small   \LaTeX-ed \today}}
%---------------------------------------------------------------------

\maketitle

%---------------------------------------------------------------------

\begin{abstract}
In 1904, the year before Einstein's seminal papers on special relativity, Austrian physicist Fritz 
Hasen\"ohrl examined the properties of blackbody radiation in a moving cavity. He calculated the 
work necessary to keep the cavity moving at a constant velocity as it fills with radiation and 
concluded that the radiation energy has associated with it an apparent mass such that $E=\frac{3}{8} mc^2$. 
In a subsequent paper, also in 1904, Hasen\"ohrl achieved the same result by computing the force 
necessary to accelerate a cavity already filled with radiation.  In early 1905, he corrected the
latter result to $E= \frac{3}{4} mc^2$.  This result, i.e., $m = \frac{4}{3} E/c^2$, 
has led many to conclude
that Hasen\"ohrl fell victim to the same ``mistake'' made by others who derived this relation 
between the mass and electrostatic energy of the electron.  Some have attributed the mistake 
to the neglect of stress in the blackbody cavity.  In this paper, Hasen\"ohrl's papers are examined from a modern, relativistic 
point of view in an attempt to understand where he went wrong.  The primary mistake in 
his first paper was, ironically, that he didn't account for the loss of mass of the blackbody 
end caps as they radiate energy into the cavity.  However, even taking this into account one 
concludes that blackbody radiation has a mass equivalent of $m = \frac{4}{3}E/c^2$ 
or $m = \frac{5}{3}E/c^2$ 
depending on whether one equates the momentum or kinetic energy of radiation to the momentum or
kinetic energy of an equivalent mass.  
In his second and third papers that deal with an accelerated cavity, Hasen\"ohrl concluded that 
the mass associated with blackbody radiation is $m = \frac{4}{3}E/c^2$, 
a result which, within the restricted context of Hasen\"ohrl's gedanken 
experiment, is actually consistent with special relativity.  (If one includes all components of the system,
including cavity stresses, then the total mass and energy of the system are, to be sure, related by $m=E/c^2$.)
Both of these problems are non-trivial and the surprising results, indeed, turn out to be relevant to the 
``$\frac{4}{3}$ problem" in classical models of the electron.
An important lesson of these analyses is that $E = mc^2$, while extremely useful, is not a ``law of physics"
in the sense that it ought not be applied indiscriminately to any extended system and, in particular, to the
subsystems from which they are comprised.  We suspect 
that similar problems have plagued attempts to model the classical electron.

 \vspace*{5mm} \noindent PACS: 03.30.+p, 01.65.+g, 03.50.De

\\ Keywords: Hasen\"ohrl, Einstein, Fermi, mass-energy 
equivalence, blackbody radiation, special relativity
\end{abstract}
%----------------------------------------------------------------------------
-
\section{Historical introduction}
\setcounter{equation}{0}\label{sec1}
%--------------------------------------------------------------------------
%\baselineskip 8mm
In 1904-5 Fritz Hasen\"ohrl published the three papers, all with the title 
``On the theory of radiation in moving bodies," for which he 
is best known ([Hasen\"ohrl1904a; Hasen\"ohrl1904b; Hasen\"ohrl1905] 
referred to as H1, H2 and H3).  They concerned  the mass equivalent of 
blackbody radiation in a moving cavity. The latter two papers appeared in the 
\emph {Annalen der Physik} and for his work Hasen\"ohrl won the Haitinger 
Prize of the Austrian Academy of Sciences. (In 1907 he succeeded Boltzmann as 
professor of theoretical physics at the University of Vienna.)  These three papers analyzed two different
\emph{gedankenexperiments} each of which demonstrated a connection between 
the energy of radiation and inertial mass.  In the first thought experiment,
he arrived at $E=\frac{3}{8} mc^2$ and in the second, $E = \frac{3}{4}mc^2$.
Hasen\"ohrl was working within the confines of an ether theory 
and, not surprisingly, these results were soon replaced by Einstein's 
quintessential $E = mc^2$.  Even so, it is interesting to ask "Where did 
Hasen\"ohrl go wrong?"

The notion that mass and energy are related originated well before Hasen\"ohrl's
and Einstein's papers.  As early as 1881, J.J. Thomson [Thomson1881] argued that the 
backreaction of the field of a charged sphere (the classical model of the electron) 
would impede its motion and result in an apparent mass increase of $(4/15)\mu e^2/a$, 
where $e$ was the charge on the sphere, $a$ its radius and $\mu$ the magnetic permeability.
Fitzgerald, Heaviside, Wein, and Abraham subsequently corrected Thompson's analysis 
and all concluded that the interaction of a moving electron with its field 
results in an apparent mass given by $m=\frac{4}{3}E/c^2$ where $E$ is the electrostatic 
energy of the stationary electron.  (For more on these early works, see Max 
Jammer's \emph {Concepts of Mass} [Jammer1951].)

All these investigations were of the relationship between the mass and the electrostatic energy
of the electron.  Hasen\"ohrl broadened the query by asking ``what is the mass equivalent
of blackbody radiation?"  Previous explanations as to why Hasen\"ohrl failed to achieve the
``correct'' result, i.e., $m = E/c^2$, are not particularly illuminating.  
For example, in his  \emph{Concepts of Mass in 
Contemporary Physics and Philosophy} [Jammer2000], Jammer, says only: ``What was probably the 
most publicized prerelativistic declaration of such a relationship [between 
inertia and energy] was made in 1904 by Fritz Hasen\"ohrl. Using Abraham's 
theory, Hasen\"ohrl showed that a cavity with perfectly reflecting walls 
behaves, if set in motion, as if it has a mass $m$ given by $m = 
8V\e_0/3c^2$, where $V$ is the volume of the cavity, $\e_0$ is the energy 
density at rest, and $c$ is the velocity of light.''  (For a more extensive 
discussion, see Boughn \& Rothman 2011.)  The overall impression is that few authors 
have made an effort to understand exactly what Hasen\"ohrl did.

In certain ways Hasen\"ohrl's thought experiments were both more bold and more 
well defined than Einstein's, which alone renders them worthy of study. A
macroscopic, extended cavity filled with blackbody radiation is certainly a more complicated system
than Einstein's point particle emitting back-to-back photons.  In addition, whereas the characteristics
of blackbody radiation and the laws governing the radiation (Maxwell's equations) 
were well known at the time, the emission process of radiation from point particles (atoms)
was not well understood.  Einstein simply conjectured that the details 
of the emission process were not relevant to his result. 

Another reason to investigate Hasen\"ohrl's thought
experiments is the apparent relation to the famous ``$\frac{4}{3}$ problem" of the self-energy of the 
electron (see \S 4).  Enrico Fermi, in fact, assumed 
that the two $\frac{4}{3}$'s were identical and devoted 
one of his earliest papers to resolving the issue [Fermi \& Pontremoli1923b].  
We initially attempted to understand Hasen\"ohrl's apparently incorrect results by reproducing  his
analyses.  This effort was frustrated by his cumbersome, pre-relativistic calculations that 
were not free from error.  The objective of the present paper is to introduce  
Hasen\"ohrl's two thought experiments and then achieve correct 
relativistic results (\S\ref{sec2} and \S\ref{sec3}), which will allow us to 
understand both the limtations and strengths of his proofs.
In the process we determine that the neglect of cavity stresses is not the primary issue and that 
Fermi's proof is apparently violated by Hasen\"ohrl's gedanken experiment (\S\ref{sec4}).

%----------------------------------------------------------------------------

\section{Hasen\"ohrl's first thought experiment}
\setcounter{equation}{0}\label{sec2}
%----------------------------------------------------------------------------

Considering the importance of blackbody radiation at the turn of the 20th century,
an investigation of the properties of blackbody radiation in a moving cavity was an
eminently reasonable undertaking.  Hasen\"ohrl considered the case of two blackbody
radiators (endcaps) at temperature T enclosed in a cylindrical cavity made of
reflecting walls (see Figure 1).  Initially the cavity is assumed to be void of any
radiation and at a time $t = 0$ the two radiators $A$ and $B$ are, in some unspecified
way, enabled to begin filling the cavity with radiation.  He assumes that the blackbody
radiators have sufficiently large heat capacity that they do not cool appreciably during this
process.  The two endcaps are presummably
held in place by stresses in the cavity walls; although, Hasen\"ohrl refers to these forces
as external ({\it au\ss en}) and treats them as such.  Whether he actually viewed the radiators 
as being held in place by forces external to the cavity or by  internal 
stresses  makes no difference to his subsequent analysis.  We choose to make this explicit by
supposing that the two encaps are actually held in place by external forces and are otherwise
free to slide back and forth inside the cavity.

\begin{figure}[htb]\label{cav1}
\vbox{\hfil\scalebox{1.0}
{\includegraphics{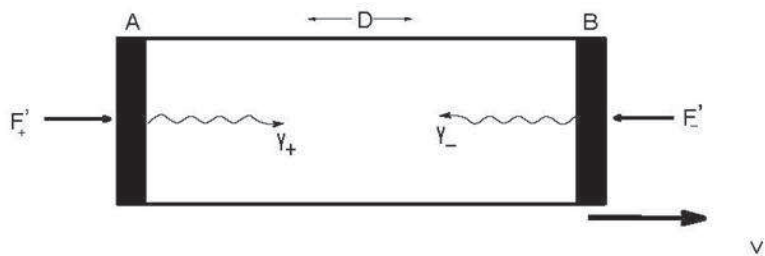}}\hfil}
{\caption{\footnotesize{A cavity consisting of two blackbody radiators, $A$ 
and $B$ in a completely reflecting enclosure of length $D$.  At a time $t=0$ 
the radiating caps suddenly begin to emit radiation in the direction of motion 
(+) and opposite the direction of motion (-).  From the frame of a moving 
observer, the (+/-) radiation will be blue/red shifted and hence exert 
different reaction forces on $A$ and $B$.  (Based on H1, H2.)}}}
\end{figure}

In the rest frame of the cavity, the radiation reaction forces on the two endcaps
are equal and opposite as are the external forces required to hold the endcaps
in place.  As viewed by a moving observer, however, the situation is quite different.
In this observer's frame, the radiation from the trailing endcap ($A$) is Doppler shifted
to the blue while radiation from the leading endcap ($B$) is redshifted.  Therefore,
when the radiators are switched on, the moving observer finds that two different external
forces $F_+$ and $F_-$ are required to counter the radiation reaction forces on the 
endcaps and keep them moving at a constant velocity.  Because the endcaps are in motion 
and because $F_+ \neq F_-$, the interesting consequence is that net work is performed 
on the cavity.

Hasen\"ohrl does not use the terminology ``rest frame of the cavity" or ``lab frame."  
Although he mentions the {\it ether} only three times throughout his papers, it is clear that 
for him all motion is taking place relative to the absolute frame of the ether. 
In this paper the lab and cavity frames have their usual meanings.
Quantities referring to the lab frame are designated by a prime; cavity-frame 
quantities are unprimed.

The crux of Hasen\"ohrl's analysis is a calculation of the work done by 
the external forces from the time that the blackbody radiators are turned on 
to the time that the cavity is at equilibrium and filled with blackbody radiation.
To order $v^2/c^2$, it turns out that Hasen\"ohrl's result, $W' = \frac{4}{3}E(v^2/c^2)$, 
is precisely the same as given by special relativity (to the same order in $v/c$).
For this reason and because  
Hasen\"ohrl's pre-relativistic calculation is very difficult to follow, we 
use a proper relativistic analysis to compute the value of the radiation 
reaction forces and the work performed by the external forces 
required to balance them.  That the two results agree is not surprising because
the reaction forces are only needed to first order in $v/c$ and can be derived from  
the non-relativistic Doppler shift and abberation relations.

%----------------------------------------------------------------------------

\subsection{Relativistic calculation of the work}
\label{subsec2.1}
%----------------------------------------------------------------------------

The strategy is to calculate the radiation 
pressure on a moving surface by transforming the blackbody radiation intensity $i$ in 
the cavity rest frame to a frame moving at velocity $v$ relative to 
the cavity. A detailed derivation of this tranformation can be found in Boughn \& Rothman[2011];
however, the same result can be got more directly from an expression
for the anisotropic temperature of the cosmic 
background radiation [Peebles \& Wilkinson1968].
Peebles and Wilkinson found that in a moving frame the radiation maintains a blackbody spectrum
\be
i'_\n = \frac{2h\n'^3}{c^2}(e^{h\n'/kT'-1})^{-1}
\ee
but with a temperature $T'$ that depends on
direction,
\be
T(\theta') = T(\frac{(1-\b^2)^\frac{1}{2}}{(1-\b\cos\th')}
\ee
where $T$ is the blackbody temperature in the rest (cavity) frame, $\b \equiv v/c$, 
and $\th'$ the angle between the radiation and $\bf{v}$.
The integral of $i'_\n$ over all frequencies is well known to be $\propto T'^4$.
Therefore the intensity in the lab frame is 
\be
i' = i \frac{(1-\b^2)^2}{(1-\b\cos\th')^4} \label{ith}
\ee
where $i$ is the intensity in the cavity frame and is given by the usual Planck formula.
Although this derivation of Eq. (\ref{ith}) assumes blackbody radiation, it is 
straightforward to show that it holds for any isotropic radiation field [Boughn \& Rothman2011].
It will become apparent that in order to calculate the work $W'$ to second order in $\b$, 
one can ignore terms of order $\b^2$ in Eq. (\ref{ith}), i.e., the relativistic corrections.   
That one need only use the non-relativistic transformation laws to compute the work at this order 
provides an explanation as to why Hasen\"ohrl obtained an essentially correct result for the work.
Finally, it is well known that the intensity and energy density of blackbody radiation (or any isotropic
radiation field for that matter) is
\be
\r = \frac{4\pi i}{c},  \label{ro}
\ee
which Hasen\"ohrl accepts (H1).  

Now consider the radiation being emitted at an 
angle $\th'$ from the left end cap of the cavity in figure 1.  Using the relation between 
momentum and energy for electromagnetic radiation $P = E/c$, the rate at which momentum
leaves that end cap is given by
\be
\frac{dP'}{dt'} = \frac{i'}{c^2} \; d\O' A \; (c\cos\th' - v),
\label{dEdt}
\ee
where $A$ is the area of end cap.  By symmetry, the only non-vanishing component 
of the momentum is in the direction of $v$.  The last factor in Eq. (\ref{dEdt}) is due to the lab frame 
relative velocity between the radiation and the encap in this direction.
From Newton's third law, $dP'/dt'$ is the magnitude of the rightward external 
force needed to counter the radiation reaction force and keep the left endcap 
moving at constant velocity.  The work done by that portion of the external force 
needed to counter the reaction force of the radiation emitted at angle $\th'$ is
\be
dW'_+ (\th') = \frac{dP'}{dt'} \cdot v \cdot \Delta t'({\th'}),
\ee
where $\Delta t'({\th'})$ is the light-crossing time for radiation at an angle $\th'$.  
After this time, radiation will be absorbed by end cap $B$ and the force necessary to
counter the resulting radiation pressure on $B$ is equal and opposite to the force on end cap $A$.
It is straightforward to show that, independent of the number of reflections along the cavity side wall,
\be
\Delta t'(\th') = \frac{D'}{c\cos\th' - v)}  \label{dt'}
\ee
where $D' = D/\g$ is the Lorentz contracted length of the cavity in the lab frame.
For cylindrical symmetry $d\O' =  2\pi\sin\th' d\th' = -2\pi d(\cos\th')$ and 
hence
\be
dW'_+ (\th') = -2\pi i \frac{A D' v (1-\b^2)\cos\th' d(\cos\th')}{c^2(1-
\b\cos\th')^4}. \label{dW+}
\ee
Retaining terms to first order in $\b$, the total work on radiator 
$A$ is
\bea
W'_+ &=& \frac{2\pi i A D v}{c^2} \int_0^1 (1+4\b\cos\th')\cos\th' 
d(cos\th')\nonumber\\
      &=&  \frac{2\pi i A D v}{c^2} \left(\frac1{2} + \frac{4\b}{3}\right).
\eea
Note that, because of abberation, the upper limit on the above integral is not precisely 
unity; however, this correction is also higher order in $\b$ and can be neglected.
The work on the right end cap can be found by taking the negative of the above expression after 
reversing the sign on $\b$.  The net work done by the external force is consequently
\be
W' = W'_+ + W'_- = \frac{4}{3}\left[\frac{4\pi i A 
D}{c}\right]\frac{v^2}{c^2}
\ee
where $AD = V$ is the rest frame volume of the cavity.
From Eq. (\ref{ro}), the quantity in brackets is 
$\r V = E$, the energy of the blackbody radiation in the cavity rest frame.  Therefore,
\be
W' = \frac{4}{3} E \frac{v^2}{c^2},
\label{W'}
\ee
which is exactly Hasen\"ohrl's result.  

One might worry that that we have ignored questions of simultaneity that, afterall, are first order in $v/c$.  
If the two endcaps begin radiating at the same time in the cavity rest frame, then in the 
moving lab frame, to first order in $v/c$, the trailing endcap will begin radiating $\delta t' = vD/c^2$ earlier.
However, the time interval $\Delta t'$ (see Eq. 2.7) used to compute the work is the lab frame time interval 
required for radiation emitted from endcap $A$ to reach endcap $B$, a quantity that is independent of when
the radiation is emitted from $A$.  The same is true for the radiation emitted from endcap $B$ and absorbed by
endcap $A$.  The only difficulty that might arise is if encap $B$ were required to absorb radiation before it 
begins to emit radiation (and the same situation for endcap $A$).  However, the shortest length of time rquired
for this to happen is $D/c$, the light travel time across the cavity, and this is much greater than $\delta t'$ 
from above.

Even so, Hasen\"ohrl's calculation is not without error.  As pointed out in the
abstract, his primary mistake was the ironic omission of the mass loss of the end caps
as they radiate energy into the cavity. Newton's second law implies that an external 
force must be applied to an object that is loosing mass if that mass is to maintain a
constant velocity.  To first order in $v/c$, it is sufficient to consider the non-relativistic
expression of the second law, i.e.,
\be
F' = dP'/dt' = d(mv)/dt' = v dm/dt' + m dv/dt' = v dm/dt'.
\ee
where $F' = F'_{ext} + F'_{rad}$, $F'_{ext}$ is the external force,
and $F'_{rad}$ is the reaction force of the radiation on the end caps.  Thus
\be
F'_{ext} = v dm/dt' - F'_{rad}
\ee
and the work due to the external force is just 
\be
W' = \int F'_{ext} v dt' = \int v^2 dm - \int v F'_{rad} dt'.
\ee
We have already computed the second term on the right.  From Eq. (\ref{W'}), it is 
just Hasen\"ohrl's $\frac{4}{3} E \b^2$.  The first term on the right is 
simply $\Delta m v^2$.  It is now necessary to use the relativistic result that
$\Delta m$ must be equal to minus
the energy lost by the end caps (divided by $c^2$), i.e., $\Delta m = -E/c^2$ where E is the energy radiated
into, and therefore the energy content of, the cavity.  Thus, the total work performed by external forces is
\be
W' = - E \frac{v^2}{c^2} + \frac{4}{3} E \frac{v^2}{c^2} = \frac{1}{3} E \frac{v^2}{c^2}.
\label{work}
\ee

Whereas Hasen\"ohrl equated the external work to the kinetic energy of the radiation in
the cavity, we must now consider the entire energy of the system, radiation plus blackbody
end caps $A$ and $B$.  We again use the relativistic result that the change of energy of the end caps is,
in the lab frame, given by $\gamma \Delta mc^2$ which to second order in $\b$, is given by
$-E(1 + 1/2 \b^2)$.\footnote{One might argue that it is inappropriate to use the relativistic results
$\Delta m = -E/c^2$ and $E' = \gamma \Delta mc^2$ in an analysis that purports to derive mass-energy equivalence.
However, the reader is reminded that the present anlaysis is, indeed, relativistic and these two relations
are known to be true for any bound, stable system by virture of the theorems of von Laue[1911] and 
Klein[1918] (see \S\ref{sec4}).}  Then conservation of energy yields
\be
W' = \frac{1}{3} E \b^2 = E'- E(1 + \frac{1}{2} \b^2) .
\ee
Finally, we define the kinetic energy of the radiation to be 
\be
(E' - E) = \frac{5}{6} E \b^2.
\label{E'-E}
\ee 
If one were to interpret this kinetic energy as due to an effective mass of the radiation, as
did Hasen\"ohrl, i.e., $ (E' - E) = \frac{1}{2}m_{eff} v^2$, then one finds that
\be
m_{eff} = \frac{5}{3} E/c^2,
\ee
This value is not Hasen\"ohrl's $\frac{8}{3} E/c^2$; however, neither is
it $E/c^2$ as one might expect from special relativity.

One might also choose to determine the effective mass from momentum conservation rather than energy
conservation.  Because the velocity is constant, this result is easily deduced from the above analysis.
The total momentum impulse to the system delivered by external forces is 
\be
\Delta P'_{ext} = \int F'_{ext} dt' = \frac{1}{v} \int v F'_{ext} dt' = W'/v = \frac{1}{3} E \frac{v}{c^2}.
\label{dpext}
\ee
The change in momentum of the end caps, to first order in $v/c$, is
$\Delta m v = - E v/c^2$.  Therefore, by conservation of momentum,
\be
\Delta P'_{ext} = \Delta m v + P'_{rad} = -(E/c^2)v + P'_{rad}
\ee
where $P'_{rad}$ is the net momentum of the radiation.  Then from Eq. (\ref{dpext})
\be
P'_{rad} = \frac{4}{3} (E/c^2) v.
\label{momentum}
\ee
  Attributing this momentum to an effective mass of the radiation,
i.e., $P'_{rad} = m_{eff} v$, implies that 
\be
m_{eff} = \frac{4}{3} E/c^2,
\ee
which is different from both of the results discussed above.  In order to make sense of all this, we
turn to the special relativistic definition of energy and momentum for radiation.

%----------------------------------------------------------------------------

\subsection{Energy-momentum tensor}
\label{subsec2.2}
%----------------------------------------------------------------------------

It is straighforward to calculate a lab frame expression for the radiative energy in the cavity 
using Eq. (\ref{ith}) and integrating over the times it takes radiation from the two end caps 
to fill the cavity (see Eq. \ref{dt'}).  The total radiative momentum in the lab frame can be 
computed in the same way,
noting that radiative momentum is radiative energy divided by the speed of light and taking into 
account the opposite directions of the momenta emitted from the two end caps.  A more direct way of
obtaining these results is simply by transforming the energy-momentum tensor of the the radiation 
from the cavity frame to the lab frame.  In addition, this formalism will be useful in our anlaysis
of Hasen\"ohrl's second gedanken experiment in \S\ref{subsec3.1} below.

The energy-momentum tensor for blackbody 
radiation (in the cavity frame) is the same as for a perfect fluid with equation of state $p=\r/3$ 
and has the form, i.e., $T^{00} = \r$, $T^{0i} = T^{i0} = 0$, $T^{ij} =p\,\delta_{ij}$, where
$\r$ and $p$ represent the energy density and pressure of the radiation in the cavity frame.
Because $T^{\m\n}$ is a tensor quantity, it is straightforward to express it in
any frame as
\be
T^{\m\n} = \frac1{c^2}(\r + p)u^\m u^\n + \h^{\m\n}p. \label{Tuv}
\ee
Here, all the symbols have their usual meanings:  $\bf u \equiv (\g c , \g 
\bf v)$ is the four velocity of the frame, $\bf v$ is the three-velocity, $\g \equiv (1 - 
\b^2)^{-1/2}$ and the metric tensor $\h^{\m\n} \equiv (-1,+1,+1,+1)$.  Greek 
indices range from 0 to 3 and Latin indices take on the values 1 to 3.  Thus, in the lab frame
\be
T'^{0 0} = (\r + p)\g^2 -p =  \r\g^2 + \frac{\r}{3}(\g^2 - 1),\label{T00}
\ee
and
\be
T'^{0 x} = (\r + p)\g^2\frac{v}{c}  = \frac{4}{3}\r\g^2\frac{v}{c}
\label{T0k}
\ee
where $x$ indicates the direction of motion which is parallel to the cavity axis.

Because $T'^{00}$ represents energy density, the total energy in the lab frame 
is
\be
E' = \int T'^{00} dV'  \label{E'1}
\ee
where $dV' = V/\g$ is the volume element in the lab frame.  
Therefore,
\be
E' = \g^{-1}T'^{00}V \label{E'2}
\ee
and from Eq. (\ref{T00}),
\be
E' =  \g E(1+ \frac{\b^2}{3}) = E(1+\frac{5}{6}\b^2) + {\cal O}(\b^4). 
\label{E'3}
\ee
This expression is the same as Eq. (\ref{E'-E}) and indicates that, to
second order in $\b$, the work $W'$ in Eq. (\ref{work}) 
is consistent with  the relativistic expression for energy in Eq. (\ref{E'1}).

Similarly, from Eq. (\ref{T0k}), the total momentum of the radiation in the 
lab frame is
\be
P' = \frac{1}{c} \int T'^{0x} dV',
\label{P}
\ee
or
\be
P' =\frac{4}{3}E\g \frac{v}{c^2} = \frac{4}{3}E \frac{v}{c^2} + {\cal 
O}(\b^3). \label{G'}
\ee
Likewise, that this expression is the same as 
Eq. (\ref{momentum}) indicates that Eq. (\ref{P}) is, indeed, the relativistic
momentum of the blackbody radiation in the lab frame.

We are left with the dilemma that there seem to be two different effective masses,
$m_{eff} = \frac{5}{3} E/c^2$ and $m_{eff} = \frac{4}{3} E/c^2$, associated with
blackbody radiation and neither of these is the expected $m_{eff} = E/c^2$.
This is a direct consequence that our definition of the total radiative
energy and momentum, $\int T'^{\m 0} dV'$, is not a covariant expression, i.e.,
$(E',P'^i)$ is not a proper 4-vector.  That the total energy/momentum of an extended
system behaves this way lies at the center of the previously mentioned
``$\frac{4}{3}$ problem'' of the self-energy of the electron.  We will return to this issue
after analyzing Hasen\"ohrl's second gedanken experiment.

%----------------------------------------------------------------
\section{The slowly accelerating cavity}
\setcounter{equation}{0}\label{sec3}
%----------------------------------------------------------------

\subsection{Hasen\"ohrl's second thought experiment}
\label{subsec3.1}

Hasen\"ohrl's first gedanken experiment, suddenly switching on two blackbody endcaps that
subsequently fill a cavity with radiation, may perhaps seem a bit contrived.  A more 
natural process would be to accelerate a cavity already filled with blackbody radiation
and this is precisely what Hasen\"ohrl considered in his second paper (H2).  On the other
hand, an accelerating blackbody cavity is a more complicated system.  In particular, one
must worry whether or not  the radiation remains in thermal equilibrium during the 
acceleration and whether or not the accelerated blackbody endcaps change their emission
properties.  Hasen\"ohrl was well aware of such problems.  He sought to mitigate them by
imagining that the process be carried out reversibly/adiabatically by requiring the 
that the velocity change happens ``infinitely slowly".  He also envisioned blackbody
endcaps with heat capacities so small that their heat contents were negligible; their
only purpose is to thermalize the radiation.  In our analysis, we obviate the problem
of thermal equilibrium by assuming the acceleration has been in effect for a very long time
so that the cavity comes to eqilibrium.  Because of the absolute frame of the ether, this
assumption wasn't availible to Hasen\"ohrl.  Even so, in our analysis we must assume that
the acceleration is small in the sense that $aD/c^2 \ll 1$.

As in his first gedanken experiment, Hasen\"ohrl computed the work required, in this case, to
accelerate the cavity to a speed $v$.  Initially, he obtained the same result as in H1, i.e.,
$W = \frac{4}{3}E\b^2$, which implied that  $m = \frac{8}{3}E/c^2$.  
After Abraham pointed out a simpler way to calculate the mass, as 
the derivative of the electromagnetic momentum with respect to velocity: $m = 
d(\frac{4}{3}E v/c^2)/dv = \frac{4}{3}E/c^2$, Hasen\"ohrl uncovered a factor of two 
error in H2, which brought him into agreement with Abraham.  He 
subsequently published the correction in paper H3. This is, perhaps, why some have concluded
that Hasen\"ohrl did nothing different from Abraham.  However, Abraham's analysis was of the
classical electron, while Hasen\"ohrl's was of blackbody radiation.

Hasen\"ohrl's calculation in H2 is extremely involved.  He did not calculate 
the work directly, but rather calculated the  small change in energy 
of the already filled cavity due to an incremental change in 
velocity.  He equated the difference between this energy and that radiated by the endcaps
to the incremental work performed on the system.  We now present a modern analysis of this 
gedanken experiment.

Suppose that the cavity is already filled with blackbody radiation
and assume that the acceleration has been applied for a sufficiently long time
that the cavity is in equilibrium.  This doesn't violate the condition that the
cavity is intially at rest in the lab frame; we simply choose the lab frame to
be the inertial frame that is instantaneously comoving with the cavity at 
$t = 0$.  We also assume that the blackbody end caps each radiate according to
Planck's law when observed in an instantaneously co-moving inertial frame.  That is,
we assume that an {\it ideal} blackbody is not affected by acceleration.  This
is analogous to the special relativistic assumption that {\it ideal} clocks are not affected by acceleration.
Of course, whether or not real blackbodies or real clocks behave this way is open to question; however,
one might expect that this is the case for very small accelerations.  In any case, 
this is our \emph{ansatz} that we will justify later.  Finally, we ignore the mass of the cavity.
One needn't assume the mass is negligible but rather only that including it doesn't
change the results of the analysis.  This will be justified shortly using the results of \S\ref{subsec3.2}.

With these assumptions, it is straightforward to demonstrate that, in an instaneously 
comoving frame, the radiation is isotropic at every point in the cavity.  This follows
directly from Liouville's theorem, i.e., phase space density is constant along 
every particle trajectory.  For photons, phase space density is proportional to $i_{\n}/\n^3$
(e.g. Misner \textit{et al.} 1973).
We assume that at the blackbody end cap, $i_{\n}$ is given by the Planck law and is,
therfore, isotropic.  Then the intensity of the radiation at a perpendicular distance
$x$ from the trailing end cap is given by
\be
i_{\n}(x) = \left(\frac{\n}{\n_e}\right)^3 i_{\n_e}
\label{L1}
\ee
where $\n_e$ indicates the frequency of the photon emitted from the trailing end cap and
$\n$ the frequency of that same photon at the point $x$.  It is straightforward to show that in the
instantaneously co-moving frame these two frequencies are related by
\be
\n \approx \n_e \left(1 - \frac{ax}{c^2}\right)
\label{L2}
\ee
where $a$ is the acceleration of the cavity.  The relation is valid regardless
of where on the end cap the photon originated.  (This is the equivalent of
gravitation redshift to which we will return in \S\ref{subsec3.2}.)  Thus
\be
i_{\n}(x) = \left(1 - \frac{ax}{c^2}\right)^3 i_{\n_e}
\label{L3}
\ee
Because $i_{\n_e}$ is given by the Planck function and therefore independent of direction,
the implication is that $i_{\n}$ is also isotropic.  Of course, 
one must consider the Doppler shifted photons emitted from the leading encap
and these photons are blue shifted.  It turns out that in order to be in thermal
equilibrium, the leading end cap must be at a lower temperature than the trailing
end cap with the result that the intensity of photons emitted from the leading 
end cap is precisely the same as that of the photons emitted from the trailing
end cap.  (This argument will be elaborated on in \S\ref{subsec3.2}.)  The result is that,
at least to first order in $ax/c^2$, the radiation in the cavity is istropic.

In this case, we can again use the perfect-fluid form of 
the stress-energy tensor Eq. (\ref{Tuv}) to describe the radiation.  
From conservation of energy/momentum we know that in any inertial frame
$T^{\m\n}\,_{,\n} = 0$ within the cavity.
The spatial part ($\m = i$) of this relation can be expressed in terms 
of the pressure, energy density and ordinary vector velocity $\bf v$ as
[Weinberg1972]
\be
\frac{\partial\mathbf{v}}{\partial t} + (\mathbf{v \cdot \nabla})\mathbf{v} =
-\frac{c^2(1 - \b^2)}{(\r + p)}\left[\mathbf{\nabla} p  + 
\frac{\mathbf{v}}{c^2} \frac{\partial p}{\partial t}\right]
\ee
where $\bf v$ is the velocity of the cavity in the lab frame.  (In all that follows, we
refrain from distinguishing primed and unprimed frames since all calculations will
be carried out in the inertial laboratory frame.)  Because the cavity
is assumed to be in equilibrium, the co-moving inertial frame pressue $p$ and density $\r$ are independent of time.
In addition, for small velocities we can discard terms that are second order in $\b^2$ and the $x$ component
of this relation becomes
\be
\frac{\partial p}{\partial x} = - \frac{(\r + p)}{c^2} = -\frac{4 p a}{c^2}.
\label{dpdx}
\ee
To first order in $ax/c^2$, the solution to Eq. (\ref{dpdx}) is 
\be
p = p_0 \left(1 - \frac{4xa}{c^2}\right) = p_0 - \frac{4}{3c^2}\r_0 a x
\ee
where $p_0$ and $\r_0$ are the radiation pressure and energy density at the trailing end of the cavity.
Finally, the forces that must be applied to the trailing and leading end caps of the cavity in order to
maintain the acceleration must be $F_+ = p_0 A$ and $F_- = -p_0 A + (4AD\r_0/3c^2)a$
where A is the area of each end cap and D is the length of the cavity.  Therefore, the total 
force on the cavity must be 
\be
F = F_+ + F_- = \frac{4AD\r_0}{3c^2}a = \frac{4}{3}\frac{E}{c^2}a
\label{netforce}
\ee
where $E$ is, to lowest order, the radiation energy in the cavity co-moving frame.

We have made several assumptions in this derivation that need justification.  First,
we assumed that $Da/c^2 << 1$.  This assumption is the requirement that the change in velocity
of the cavity in one light crossing time is much less than the speed of light, the small
acceleration condition.  We have neglected any change in $p$ and $\r$ in the transverse 
directions.  The mirrored sidewall of the cylindrical cavity has the same effect on the radiation in
the cavity as encaps of infinite transverse extent, in which case pressure and density 
only depend on $x$.  The approximation that
$E = \r_0 AD$ neglects terms of order $Da/c^2$ but these only change the total force by terms
that are second order in this quantity.  In addition, we did not address what constitutes a 
constant acceleration of the cavity.  Hasen\"ohrl's pre-relativistic scenario assumed a \emph{rigid} cavity.
In our calculation, we interpret the constant acceleration to be such that the cavity is \emph{Born rigid},
that is, the cavity remains the same length in all instaneous co-moving intertial frames
as would be expected for a cavity in an equilibrium state.  Born rigid acceleration requires that
the acceleration $a_l$ of the leading end cap is related to the acceleration $a_t$ of the 
trailing end cap by [Newman \& Janis1959]
\be
a_l = \frac{a_t}{1 + a_tD/c^2}.
\ee
Therefore, the approximation that $a_l \approx a_t \approx a$ again neglects terms of order
$Da/c^2$ and only changes the net force by terms second order in this quantity.  Finally, we show in 
\S\ref{subsec3.2} that the fractional changes in temperature and pressure of blackbody radiation in an accelerated 
frame are of order $aL/c^2$ where $L$ is the relevant dimension of the system in the direction of the
acceleration.  Therefore, one might expect that the deviation of a blackbody radiator (or an 
electromagnetic clock for that matter) in an accelerated frame would be of order $aL/c^2$
where $L$ is some characteristic length of the process.  For a blackbody
radiator (or atomic clock) L might be the size of an atom, or the mean free length of a photon within 
the blackbody absorber, or perhaps the wavelength of the radiation.  In any case, because $L$ is much, much
smaller than the size of the cavity, such effects are negligible. This provides justification for
assuming that the blackbody end caps do, indeed, radiate according to Planck's Law.

If we identify the effective mass of the radiation in terms of $F = m_{eff} a$, then 
Eq. (\ref{netforce}) implies that
\be
m_{eff} = \frac{4}{3} E/c^2
\label{meff2}
\ee
in agreement with our momentum analysis of Hasen\"ohrl's first gedanken experiment in \S\ref{subsec2.1} and with
what Hasen\"ohrl found in his second gedanken experiment albeit using a conservation of energy argument.
We can reproduce his energy argument result simply by integrating the net force in Eq. (\ref{netforce})
\be
W = \int F v dt = \frac{4}{3}\frac{E}{c^2} \int a v dt = \frac{2}{3}E\b^2,
\ee
which is precisely what Hasen\"ohrl found.  Upon equating this work with kinetic energy of the radiation
expressed as $m_{eff} = 1/2 m v^2$, he found that the effective mass was that given by Eq. (\ref{meff2}).
On the other hand, our work/energy analysis of H1 found that $m_{eff} = \frac{5}{3} E/c^2$.
Where have we (and Hasen\"ohrl) gone wrong?  

Hasen\"ohrl was certainly familiar with Lorentz-Fitzgerald contraction and, in fact, invoked it in H2 and H3, although,
not in his calculation of the work performed by the external forces.  Because a Born rigid object has constant
dimensions in instanteously co-moving frames, its length in the lab frame is Lorentz contracted.  This 
is only approximately so.  Because of their different accelerations, the velocities of the two ends of the
cavity are not the same in the lab frame.  Never the less, the usual expression for Lorentz contraction is
valid to second order in $\b$.  Therefore, the distance moved by the leading end cap is less than that moved
by the trailing end cap by an amount, $D - D/\gamma \approx \frac{1}{2}D \b^2$.  The work performed by the
external forces in accelerating the cavity from rest to a velocity $v$ is then
\be
W = \int F_+ v_+ dt + \int F_- v_- dt =\int F_+ \Delta x_+ + \int F_- \Delta x_-
\ee
where $\Delta x_+ - \Delta x_- \approx \frac{1}{2}D\b^2$.  Substituting the expression for $F_+$ and 
$F_-$ from above we find,
\be
W = \frac{4AD\r_0}{3c^2}a \Delta x_- + Ap_0 (\Delta x_+ - \Delta x_-) = 
\frac{4AD\r_0}{3c^2}a \Delta x_- + Ap_0 \frac{1}{2} D \b^2.
\ee
Now $\Delta x_+ = \Delta x_- + O(\b^2)$ and the displacement $\Delta x$ is related to the velocity $v$ and
acceleration $a$ by $v^2 \approx 2 a \Delta x$.  Also $A p_0 D \approx E/3$ and $A D \r_0 \approx E$.  Thus
the net work performed by the forces in the lab frame is
\be
W = \frac{5}{6}E\b^2.
\ee
Setting this equal to $\frac{1}{2} m_{eff} v^2$ gives $m_{eff} = \frac{5}{3}E/c^2$, precisely the same result as we
got from our conservation of energy analysis of Hasen\"ohrl's first gedanken experiment.

So it seems that a proper analysis of Hasen\"ohrl's two gedanken experiments give consistent results and are also
consistent with the relatistic expressions for energy and momentum of blackbody radiation.  The problem is that the
results from energy conservation imply an effective mass that is different from that implied by conservation of
momentum and both of these are different from the $m_{eff} = E/c^2$ that we are led to expect from special
relativity.  This dilema is closely associated with a similar situation for classical models of the electron and
we return to these issues in \S\ref{sec4}.  First, however, we consider an analogous situation of 
a blackbody cavity at rest in a uniform, static gravitational field.

%XXXXXXXXXXXXXXXXXXXXXXXXXXXXXXXXXX
\subsection{Blackbody cavity in a static gravitational field}
\label{subsec3.2}
%XXXXXXXXXXXXXXXXXXXXXXXXXXXXXXXX

Suppose the cylindrical blackbody cavity is at rest in a static, uniform gravitational field with the axis of the cavity in the direction 
of the field.  We again use Liouville's theorem, Eq. (\ref{L1}), this time in combination with the usual equation for the
gravitational redshift of photons, i.e.,
\be
\n \approx \n_e \left(1 - \frac{gx}{c^2}\right)
\label{g1}
\ee
where $\n_e$ is the frequency of a photon emitted from the bottom end cap, $\n$ is the frequency of that same photon at
a height $x$ above the bottom end cap, and $g$ is the local acceleration of gravity.  
Of course, by the equivalence principle, this expression is the same as Eq. (\ref{L2}) with $g = a$.  Combining 
Eqs. (\ref{L1}) and (\ref{g1}) again yields Eq. (\ref{L3}) with $a = g$, i.e., the upward intensity of radiation at
point $x$ due to photons emitted from the bottom end cap.
The upward directed intensity of the radiation incident on the top end cap $i_{\n}(D)$ due to the intensity of
radiation emitted by the bottom end cap $i_{\n_e}(0)$ is given by
\be
i_{\n}(D) = \left(1 - \frac{gD}{c^2}\right)^3 i_{\n_e}(0).
\label{g2}
\ee
The total flux incident on the upper end cap is 
\be
f_u = \int \int i_{\n}(D) d\n cos(\theta) d\Omega = \left(1 - \frac{gD}{c^2}\right)^4 \int i_{\n_e}(0) d\n_e \int cos(\theta) d\Omega.
\ee
The integrals on the right hand side of this equation are well known and their product is given by 
$\sigma T(0)^4$ where $T(0)$ is the temperature of the lower end cap and $\sigma$ is the Stefan-Boltzmann constant.  
On the other hand, the flux emitted by 
the upper blackbody end cap is the usual $\sigma T(D)^4$.  These
two fluxes must be equal if the system is in equilibrium, thus $T(D) = (1 - gD/c^2) T(0)$.
Using this relation, it is straightforward to show that the downward intensity at an 
interior point $x$ due to photons emitted 
from the upper end cap is equal to the upward intensity of the photons emitted from the 
lower end cap at the same point, as 
was asumed in \S\ref{subsec3.1}.

In fact, it is easily demonstrated that the radiation at any point $x$ in the 
interior of the cavity has a blackbody spectrum 
characterized by a temperature $T(x) = (1 - gx/c^2) T(0)$.  
From the Planck formula we know that the phase space density of blackbody radiation is 
\be
\frac{i_{\n}}{\n^3} \propto \left(e^{\frac{h\n}{kT}}-1\right)^{-1}.
\ee
We assume that the radiation emitted from the bottom end cap has a blackbody spectrum and, therefore, obeys this relation.
By Liouville's theorem, the phase space density of the radiation at point $x$, is equal to that of the emitted radiation,
i.e., 
\be
\frac{i_{\n}}{\n^3} = \frac{i_{\n_e}}{\n_e^3} \propto \left(e^{\frac{h\n_e}{kT(0)}}-1\right)^{-1} 
= \left(e^{\frac{h\n}{kT(x)}}-1\right)^{-1}
\ee
where from Eq. (\ref{g1}) $T(x) \equiv (1 - gx/c^2) T(0)$.  Therefore, the radiation at $x$ also has a blackbody 
spectrum with a characteristic temperature $T(x)$.  One can easily demonstrate that the same result is obtained by
considering the gravitational blueshifted photons emitted from the top end cap.

Now imagine that the top and bottom end caps of the cavity are held in place not by internal cavity stresses but rather 
by external forces, i.e., the end caps are otherwise free to slide up and down inside the cavity.  
The  radiation pressure pushing down on the lower end cap is $p(0) = \r(0)/3 \propto T(0)^4$
while the pressue pushing up on the upper end cap is $p(D) = \r(D)/3 \propto T(D)^4$. 
Therefore, $p(D) = (1 - aD/c^2)^4 p(0)$.  The force required to
support the bottom end cap is its weight, $M_{ec} g$, plus the force required 
to balance the pressure, $p(0)A$ and the force required to support the top end cap is clearly $M_{ec}g - p(D)A$.  In
addition, of course, a force $M_{sw} g$ is needed to support the side wall of the cavity. 
Finally the total force required to support the entire cavity, including radiation, is 
\be
F \approx M g + \frac{4ADp(0)}{c^2}g \approx \left(M + \frac{4}{3} \frac{E}{c^2}\right) g
\label{gravity}
\ee
where $M = 2M_{ec} +M_{sw}$ is the total mass of the cavity.  It is clear from this expression that the 
weight of the radiation, $m_{eff} g$, implies an effective mass of $\frac{4}{3} E/c^2$, 
the same as Hasen\"ohrl deduced and consistent with the results of our momentum analyses for both of 
Hasen\"ohrl's gedanken experiments.  In the gravitational case there is no work performed by 
the external forces and, hence, no analog of our work/energy analyses.  Eq. (\ref{gravity}) also justifies neglecting the
mass of the cavity in \S\ref{subsec3.1}.  Again, we find a result that seems to contradict Einstein's $E = mc^2$.

This seeming contradiction and the connection with similar results for the classical electron brings us to a more general
discussion of the energy and momentum of extended objects.

%%%%%%%%%%%%%%%%%%%%%%%%%%%%%%%%%%%%%%%%%%%%%
\section{Hasen\"ohrl, Fermi, and the classical model of the electron}
\setcounter{equation}{0}\label{sec4}
%%%%%%%%%%%%%%%%%%%%%%%%%%%%%%%%%%%%%%%%

In \S\ref{sec2} and \S\ref{sec3} we found momentum convservation and energy conservation in Hasen\"ohrl's two
gedanken experiments led to two different effective masses associated with blackbody radiation, 
$m_{eff} = \frac{4}{3} E/c^2$ and $m_{eff} = \frac{5}{3} E/c^2$.  Futhermore,
these two masses were found to agree with the standard expressions for energy and momentum , i.e.,
$E = \int T^{00} dV$ and $P^i = \frac{1}{c} \int T^{0i} dV$.  That these two expressions lead to different effective masses
is a direct consequence of the integrals not being Lorentz covariant, i.e., $E$ and $\bf P$ do
not constitute a covariant 4-vector.  If they did, it is straightforward to show that the expressions for both 
energy and momentum would imply an effective mass of $E_0/c^2$, the Einstein relation.    Suppose $(E,\bf P)$
is an energy/momentum 4-vector.  In the zero momentum frame this is $(E_0,0)$.  A Lorentz boost to a frame with velocity
$-\bf{v}$ immediately gives
\be
E = \gamma E_0 \approx E_0 + \frac{1}{2} E_0 \b^2 = E_0 + \frac{1}{2} \frac{E_0}{c^2} v^2
\ee
and
\be
{\bf P} = \gamma E_0 \frac{\bf v}{c^2}  \approx \frac{E_0}{c^2} \bf v.
\ee
If one identifies the kinetic energy, $E - E_0$, with $\frac{1}{2} m_{eff} v^2$ and the momentum with $m_{eff} \bf v$,
both of these relations imply $m_{eff} = E_0/c^2$.  
A solution to the delimma might be simply to redefine the total energy and momentum of an extended body, in this
case blackbody radiation, so that they are the components of a 4-vector.  On the otherhand, the two
Hasen\"ohrl gedanken experiments present the same dilemma and these results are derived from the work/energy
theorem and conservation of momentum, neither of which seems amenable to redefinition.

A similar situation occurs in the case for the energy and momentum of the electromagnetic field surrounding a 
charged spherical shell (the classical electron).  It is straightforward to show that the integral expressions 
for energy and momentum give precisely the same two results as those for blackbody radiation, i.e., Eqs. \ref{E'3} and \ref{G'}.  
Perhaps because most analyses make use of Newton's 2nd law/momentum conservation, 
historically such analyses deduced that $m_{eff} = \frac{4}{3} E/c^2$, hence, the ``$\frac{4}{3}$ problem''. 
One of the controversial issues is whether or not one must take into account the forces needed to make stable
the repulsive charge of the electron.  Poincar\'e (1906) was the first to consider the stability of the electron and
introduced ``Poincar\'e stresses," which were unidentified nonelectromagnetic stresses meant to bind the 
electron together.  With the inclusion of these stresses, one finds that the effective mass of the electron is,
indeed, $m_{eff} = E/c^2$ if one includes in $E$ the contribution of Poincar\'e stresses.
(Poincar\'e suggested more than one model for stabilizing stress[Cuvaj1968].)

Max von Laue (1911) was the first to generalize this conclusion.  He demonstrated that for any closed, static 
(extended) system for which energy and momentum are conserved, i.e., $T^{\m\n}\,_{,\n} = 0$, the energy and
momentum computed according to 
\be
P^\m = \int T^{0\m} dV
\label{ener_mom}
\ee
do indeed comprise a 4-vector.  Felix Klein (1918) extended Laue's proof to time-dependent, closed systems.
The conclusion is that for any closed, conservative system the \emph{total} energy/momentum, defined by
Eq. (\ref{ener_mom}), is a 4-vector and, as a consequence, $m_{eff} = E_0/c^2$.  (For a simple version of Klein's
proof, see [Ohanian 2012].)  As a consequence of Klein's theorem, it follows that
the 4-momentum $P^\m$ is related to the 4-velocity $u^\m$ of the zero momentum frame center of mass 
(center of energy) by $P^\m = (E_0/c^2) u^\m$ (e.g., [M\o ller1972]).  It is then straightfoward to show that, 
for any time-dependent, closed system, ${\bf F} = \gamma (E_0/c^2) {\bf a}$ where $E_0$ is the 
total energy in the zero momentum frame, $\bf a$ is the acceleration of the zero momentum frame center
of mass, and $\bf F$ is the external force on the otherwise conservative 
system.  

At first blush, the theorems of Laue and Klein
might seem to contradict our results for Hasen\"ohrl's two gedanken experiments; however, neither of these
satisfy the Laue/Klein assumption that the system is closed.  For the Hasen\"ohrl scenarios, external forces (not
included in $T^{\m\n}$) are necessary to contain the radiation.  If instead, the radiation is contained by
stresses in the cavity walls {\it and} these stresses are included in $T^{\m\n}$, then it is straightforward to show that
the total energy and momentum from Eq. (\ref{ener_mom}) are consistent with $m_{eff} = E/c^2$ where
$E$ is the total energy of the radiation plus cavity. Hasen\"ohrl certainly supposed that the radiation was
contained by the cavity; however, he chose to consider the forces due to cavity stresses as \emph{external}.
This is a legitimate and understandable point of view.  After all, Hasen\"ohrl was interested in the inertial
mass of blackbody radiation, not the combined inertial masses of the radiation plus cavity.

In two of his earliest papers, Fermi (1922 \& 1923a) took another approach to solving the $\frac{4}{3}$ problem,
one that made no mention of the Poincar\'e stresses necessary to stabilize the electron.  Fermi maintained
that the $\frac{4}{3}$ problem for the classical electron arises because the electron is assumed to be a rigid body, in 
contradiction to the principles of special relativity.  He applied the 
concept of ``Born rigidity" to the electron, which requires that given points 
in an object always maintain the same separation in a sequence of inertial 
frames co-moving with the electron. Equivalently, Born rigidity demands that 
the worldline of each point in the electron should be orthogonal (in the 
Lorentzian sense) to constant-time hypersurfaces in the co-moving frames 
(see, eg., Pauli 1921).  However, such constant-time hypersurfaces are of 
course not parallel to those in the lab.  A constant-time integration over the electron's volume in 
its rest frame assumes that two points on the electron's diameter cross the 
$t=0$ spatial hypersurface simultaneously, but this will \emph{not} be the 
case in a Lorentz-boosted frame [Boughn \& Rothman2011].  Fermi chose to evaluate the action by integrating 
over the volume contained within the constant-time hypersurfaces in the co-moving frame (equivalent to using 
Fermi normal coordinates, which he developed in an earlier paper [Fermi1923a]).  In a sense, this choice 
renders the analysis covariant, i.e., independent of the lab frame, and it is, perhaps, not surprising that
the result of his analysis is that ${\bf F} = (E/c^2){\bf a}$.  The details of Fermi's approach can be found in
Jackson [1975] and Bini [2011].  It should be emphasized that Fermi's solution to the $\frac{4}{3}$ problem,
unlike the Poincar\'e/Laue/Klein approach, is silent on any non-electromagnetic forces that hold the 
electron together.  Like Fermi, Fritz Rohrlich also sought to solve the $\frac{4}{3}$ problem without 
addressing the stability of the electron.
Rohrlich [1960] simply redefined the expression for total energy/momentum in Eq. (\ref{ener_mom}) so 
that it is covariant and, thus, constitutes a proper 4-vector.

In a second 1923 paper[Fermi \& Pontremoli1923b], Fermi and Pontremoli applied the above
prescription  to solve Hasen\"ohrl's cavity-radiation problem.  They considered 
the forces applied to a volume of radiation and restricted their attention to the slowly 
accelerated case.  Therefore, their results apply
to Hasen\"ohrl's second gedanken experiment.  
They concluded that the acceleration of the radiation in the cavity requires a force
${\bf F} = (E/c^2){\bf a}$ independent of any forces (e.g., cavity stresses) that contain the radiation.
While it might seem that Fermi's and Rohlich's insistance on a covariant approach 
is a reasonable demand, the resulting analyses do not seem to be capable of capturing the physics of a
Hasen\"ohrl-type problem.  This should give one pause.

Whether the Fermi/Rohrlich approach or that of Poincar\'e, von Laue, and Klein is the appropriate description
of the classical electron remains a controversial subject and continues to foster arguments on both
sides of the issues.  A sample over the last 50 years includes papers by: 
Rohrlich [Rohrlich1960; Rohrlich1982]; Gamba [Gamba1967]l; Boyer [Boyer1982]; 
Campos and Jim\'enez {Campos \& Jim\'enez1986]; Campos [Campos \emph{et al.}2008]; and Bini \emph{et al.}
[Bini2011].
The second edition of Jackson's \emph{Classical Electrodynamics} [Jackson1975] 
discusses both approaches.  The interested reader is 
referred to these works.  With regard to classical models of the electron, both methods
give the same result and the electron is, in any case, fundamentally a 
quantum phenomenon.

On the other hand, these issues are not ambiguous in the case of Hasen\"ohrl's 
blackbody cavity.  In this case, neither of the approaches of
the two schools is particularly helpful.  The Laue/Klein theorem cannot be invoked
because the system is not closed; the forces that contain the radiation are external
to the system.  We suspect that that members of the Laue school would agree with
this point of view. (Of course, if a blackbody cavity is stabilized by stresses within the
cavity walls, then the Laue/Klein theorem would indeed apply with the result that 
${\bf F} = (E/c^2){\bf a}$ where $E$ is the total rest frame energy of the radiation 
{\it and} cavity.)  On the other hand, Fermi's own anlaysis of
Hasen\"ohrl's slowly accelerating blackbody cavity yields a result in conflict with our
relativistic analysis.  One suspects that precisely the same would be true for a
macroscopic charged spherical shell with the charge held in place by external forces. (We
plan to analyze this system elsewhere.)  

We refrain from taking a 
point of view on the controversy regarding the structure of the fundamentally quantum
mechanical electron nor even will we argue that the Fermi/Rohrlich definition of relativistic
energy/momentum is invalid.  On the contrary, its covariant nature has a certain
appeal.  However, it is clear that the ideological application of this notion without regard
to the details of a system can lead one astray.  In particular, identifying $E/c^2$
with the effective mass of blackbody radiation leads immediately to 
${\bf F} = (E/c^2){\bf a}$ for Hasen\"ohrl's slowly accelerating cavity, which is in conflict
with a proper relativistic analysis.  One might argue that systems bound by external
forces rarely occur in problems dealing with relativistic mechanics.  This may be true;
however, the purpose of Hasen\"ohrl's gedanken experiment, and Fermi's response for that
matter, was to answer foundational problems in physics.  In this sense Hasen\"ohrl's
$m_{eff} = \frac{4}{3} E/c^2$ is correct and Fermi's $m_{eff} = E/c^2$
appears to be wrong.  

One might argue that Fermi's analysis, while not explicitly including
the forces necessary to contain the radiation, might finesse the
problem by assuming Born rigidity.  On the other hand, our relativistic analysis also assumes Born
rigidity and yet arrives at a different result.  Another possibility is that
Fermi's analysis somehow only includes that part of the external force
necessary to accelerate the radiation and ignores that part of the
force that stabilizes the cavity; however, how one might effect such
a separation of forces is not immediately obvious.  Of course, it is possible
that Fermi simply misunderstood what Hasen\"ohrl meant by ``external
forces''.  Perhaps the important lesson of this exercise is that while $E =
mc^2$ is a ubiquitous and very valuable relation, it is not a ``law of
physics'' that can be used indiscriminately without regard to the
details of the system to which it is applied.

It is often claimed that Einstein's derivation of $E = mc^2$ was the first generic proof of 
the equivalence of mass and energy (see Ohanian[2009] 
for arguments to the contrary).  It is true that 
Hasen\"ohrl's analysis was restricted to the inertial mass of blackbody radiation;
however, Einstein's  gedanken experiment involves radiation 
emitted from a point mass and, futhermore, gives no indication how this occurs.  
If it is radiation due to radioactive decay, as Einstein implies at the end of his paper, 
then perhaps it is necessary to take into account the details of this process. 
In any case, Einstein is clearly speaking about 
electromagnetic radiation, and so it  is difficult to conclude that his 
thought experiment should be taken as a general theorem about mass and energy. Einstein's
great contribution was, perhaps, that based on his simple gedanken experiment, 
he \emph{conjectured} that $E = mc^2$ was broadly true for all interactions.
Over time, his conjecture was justified theoretically and verified experimentally, 
but this was through the efforts of many scientists and engineers.

Fritz Hasen\"ohrl attempted a legitimate thought experiment and his analysis, though
hampered by a pre-relativistic world view, was certainly recognized as significant at the time.  
Whether or not his analysis was completely consistent, one of his conclusions, that the acceleration
of blackbody radiation by external forces satisfies ${\bf F} = \frac{4}{3}(E/c^2){\bf a}$,
was correct, even if of limited applicability, and for this he should be given credit.  
In addition, his gedanken experiment raises similar questions for the classical electron, 
issues that remain of interest today. Hasen\"ohrl's gedanken experiments are
worthy of study and are capable of revealing yet another of the seemingly endless
reservoir of the fascinating consequences of special relativity. \\

{\bf Acknowledgements}
I thank Tony Rothman for introducing me to this fascinating problem and for contributing
much to this paper.  Our previous joint work [Boughn \& Rothman2011] includes some
of the content of the present paper and, in addition, discusses much more of the history
of $E = mc^2$.  I am grateful to Bob Jantzen for sharing [Bini2011] before publication, and I thank him as well 
for a helpful conversation.  Thanks are due Jim Peebles for helpful 
conversations and expecially for pointing out the relevance of Liouville's theorem, and to
Hans Ohanian for criticisms that prompted a substantial change in the analysis of 
Hasen\"ohrl's second gedanken experiment.\\

{\large \bf References}

Bini, D.,  Geralico, A., Jantzen, R. and Ruffini, R. 2011.  On Fermi's 
resolution of the `$\frac{4}{3}$ problem' in the classical theory of the electron: 
hidden in plain sight. To appear in  \emph{Fermi and Astrophysics}, edited by 
R. Ruffini and D. Boccaletti.  World Scientific, Singapore, 2011.

Boyer, Timothy.  1982.  Classical model of the electron and the definition of 
electromagnetic field momentum.  \emph{Phys. Rev.} {\bf D25}: 3246--3250

Boughn, S.and Rothman, T. 2011.  Hasen\"ohrl and the Equivalence of Mass and Energy.
arXiv:1108.2250

Campos, I. and Jim\'enez, J. 1986.   Comment on the $\frac{4}{3}$ problem in the 
electromagnetic mass and the Boyer-Rohrlich controversy.  \emph{Phys. Rev.} 
{\bf D33}: 607--610

Campos, I., Jim\'enez, J. and Roa-Neri, J.  2008. Comment on ``The 
electromagnetic mass in the
Born-Infeld theory''. \emph{Eur. J. Phys.} {\bf 29}: L7–-L11

Cuvaj, Camillo.  1968.  Henri Poincar\'e's Mathematical Contributions to 
Relativity
and the Poincar\'e Stresses. \emph{Am. J. Phys.} {\bf 36}: 1102-1113

Newman, E., \& Janis, A. 1959.  Ericksen, E. et al. 1982.  Rigid Frames in Relativity.Relativistic rigid motion in one dimension.  
\emph{Phys. Rev.} {\bf 116}: 1610--1614

Fermi, Enrico.  1922.  Correzione di una contraddizione tra la teoria 
elettrodinamica e quella relativistica delle masse elettromenetiche.  
\emph{Nuovo Cimento} {\bf 25}:  159-170. English translation to appear as 
``Correction of a contradiction between the electrodynamic theory and 
relativistic theory of electromagnetic masses," in \emph{Fermi and 
Astrophysics}, edited by R. Ruffini and D. Boccaletti. World Scientific, 
Singapore, 2012

Fermi, Enrico.  1923a.  Sopra i fenomena che avvengono in vicinanza di una 
linea oraria.  \emph{Rend. Lincei} {\bf 31}: 21--23.   English translation to 
appear as ``On phenomena occuring close to a world line," in \emph{Fermi and 
Astrophysics}, edited by R. Ruffini and D. Boccaletti.  World Scientific, 
Singapore, 2012

Fermi, Enrico and Pontremoli, Aldo.  1923b.  Sulla mass della radiazione in 
uno spazio vuoto.  \emph{Rend. Lincei} {\bf 32}: 162-164. English translation 
to appear as ``On the mass of radiation in an empty space," in  \emph{Fermi 
and Astrophysics}, edited by R. Ruffini and D. Boccaletti. World Scientific, 
Singapore, 2012

Gamba, A.  1967.  Physical quantities in different reference systems 
according to relativity. \emph{Am. J. Phys} {\bf 35}: 83--89

Hasen\"ohrl, Fritz.  1904a.  Zur Theorie der Strahlung in bewegten K\"orpern.   
\emph{Wiener Sitzungsberichte} {\bf 113}: 1039-1055

Hasen\"ohrl, Fritz.  1904b.  Zur Theorie der Strahlung in bewegten 
K\"orpern," \emph{Annalen der Physik} {\bf 320}: 344--370

Hasen\"ohrl, Fritz.  1905.  Zur Theorie der Strahlung in bewegten K\"orpern, 
Berichtigung.  \emph{Annalen der Physik} {\bf 321}: 589--592

Hasen\"ohrl, Fritz.  1907, 1908.  Zur Thermodynamik bewegter Systeme. 
\emph{Wiener Sitzungsberichte} {\bf 116}, IIa (9): 1391-1405  and {\bf 117}, 
IIa (2): 207--215

Jackson, John.  1975.  \emph{Classical Electrodynamics}, second edn. John 
Wiley and Sons, New York

Jammer, Max.  1951.  \emph{Concepts of Mass}. Harvard University Press, 
Cambridge

Jammer, Max.  2000.  \emph {Concepts of Mass in Contemporary Physics and 
Philosophy}, pp. 72--73. Princeton University Press, Princeton

Klein, Felix.  1918.  \"Uber die Integralform der Erhaltungss\"atze und der 
Theorie die r\"aumlich-geschlossenen Welt. \emph{Nach. Gesell. Wissensch. 
G\"ottingen, Math.-Physik, Klasse}, 394--423

Laue, Max. 1911. \emph{Das Relativit\"atsprinzip}. Vieweg, Braunschweig

Misner, Charles, Thorne, Kip and Wheeler, John. 1973. \emph{Gravitation}. 
W.H. Freeman, New York

M\o ller, Christian.  1972.  \emph{The Theory of Relativity}. Oxford 
University Press, Oxford

Ohanian, Hans. 2009 \emph{Did Einstein Prove $E=mc^2$?}. \emph{Studies in History 
and Philosophy of Modern Physics} {\bf 40}: 167-173

Ohanian, Hans. 2012 \emph{Klein's Theorem and the Proof of $E = mc^2$}.
\emph{Am. J. Phys} in press

Pauli, Wolfgang.  1921.  \emph{Theory of Relativity}. Pergamon Press, London, 
1958

Peebles, James and Wilkinson, David.  1968.  Comment on the anisotropy of the 
primeval fireball.  \emph{Physical Review} {\bf 174}: 2168

Poincar\'e, Henri.  1906.  Sur la dynamic de l'electron. \emph{Rendiconti del 
Circolo matematico di Palermo} {\bf 21}: 129–176

Rohrlich, Fritz.  1960.   Self-energy and stability of the classical 
electron.  \emph{Am. J. Phys} {\bf 28}: 639--643

Rohrlich, Fritz.  1982.  Comment on the preceeding paper by T. H. Boyer.  
\emph{Phys. Rev.} {\bf D25}: 3251-3255

Thomson, Joseph J. 1881. On the electric and magnetic effects produced by the 
motion of electrified bodies. \emph{Philosophical Magazine}, Series 5, {\bf 
11}: 229--249

Weinberg, Steven, 1972. \emph{Gravitation and Cosmology}. 
John Wiley \& Sons, New York

\end{document}